# Sound non-reciprocity based on synthetic magnetism

Zhaoxian Chen[1,2], Zhengwei Li[1], Jingkai Weng[1], Bin Liang[1,†], Yanqing Lu[2,†], Jianchun Cheng[1,†] and Andrea Alù[3,4,†]

[1]Collaborative Innovation Center of Advanced Microstructures and Key Laboratory of Modern Acoustics, MOE, Institute of Acoustics, Department of Physics, Nanjing University, Nanjing 210093, People's Republic of China

[2]College of Engineering and Applied Sciences, Nanjing University, Nanjing 210093, People's Republic of China

[3]Photonics Initiative, Advanced Science Research Center, City University of New York, New York, NY 10031, USA

[4]Physics Program, Graduate Center, City University of New York, New York, NY 10016, USA

†Correspondence and requests for materials should be addressed to:
A. A. (email: aalu@gc.cuny.edu), B. L. (email: liangbin@nju.edu.cn),
Y. L. (email: yqlu@nju.edu.cn), J. C. (email: jccheng@nju.edu.cn)

**Abstract:** Synthetic magnetism has been recently realized using spatiotemporal modulation patterns, producing non-reciprocal steering of charge-neutral particles such as photons and phonons. Here, we design and experimentally demonstrate a non-reciprocal acoustic system composed of three compact cavities interlinked with both dynamic and static couplings, in which phase-correlated modulations induce a synthetic magnetic flux that breaks time-reversal symmetry. Within the rotating wave approximation, the transport properties of the system are controlled to efficiently realize large non-reciprocal acoustic transport. By optimizing the coupling strengths and modulation phases, we achieve frequency-preserved unidirectional transport with 45-dB isolation ratio and 0.85 forward transmission. Our results open to the realization of acoustic nonreciprocal technologies with high efficiency and large isolation, and offer a route towards Floquet topological insulators for sound.

**Introduction**

In conventional linear and time-independent media, wave propagation is usually reciprocal and the scattering matrix of the system is symmetric [1,2]. Breaking reciprocity is of fundamental interest for wave physics and it may enhance existing technologies built on the assumption of reciprocal wave transmission. In particular, the intrinsic lack of strong magneto-acoustic effects prohibits non-reciprocal steering of acoustic waves using magnetic materials as commonly done in the electromagnetic domain. Non-reciprocal steering for sound have been realized using asymmetric nonlinearities [3-5], then followed by using biased media flow [6,7]. However, non-reciprocity with nonlinear media is amplitude-dependent, and the output suffers from phase noise and frequency distortion. The introduction of directional media flow provides the possibility of breaking time-reversal symmetry without affecting the signal frequency, however such mechanisms typically introduce undesired noise at high flow speed and may be impractical in integrated acoustic technologies.

Recently, time modulation has attracted rapidly-growing attention as an effective alternative to realize non-reciprocity for photonics [8,9] and acoustics [1,10]. A synthetic linear momentum of the medium can be imparted by modulating it in time and space with a travelling wave-like modulation scheme, realizing asymmetric indirect mode transitions that break reciprocity [11-13]. Despite considerable efforts dedicated to implement such modulation with airborne sound [14-16] and elastic waves [17-20], the achieved isolation ratio has so far been limited [21,22]. On the other hand, when the medium is modulated uniformly, direct mode transitions cannot directly break reciprocity, but instead they can introduce an effective gauge field [23-25]. By combining two or more of such mode transitions, we can then realize non-reciprocity based on the introduction of a synthetic magnetic field [26-29]. However, due to the lack of efficient modulation techniques, acoustic non-reciprocity based on synthetic magnetism has been rarely reported [30,31]. To date, it still remains challenging to realize acoustic nonreciprocity with high forward transmission, large isolation, linear response and avoiding frequency conversion.

In this work, we propose to overcome these limitations by relying on synthetic magnetism imparted by time modulation, and demonstrate both theoretically and experimentally an implementation by designing an acoustic cavity system with large and efficient unidirectional transmission features. A novel mechanism to modulate the acoustic coupling both in sign and strength is introduced to achieve this goal. By controlling the coupling phases, we are able to synthesize an effective magnetic bias for sound that breaks time-reversal symmetry and realizes an efficient and frequency-preserving non-reciprocal acoustic system. The measured isolation ratio is as high as 45 dB, and the forward transmission is 0.85, which can be further improved by optimizing the design. Without relying on airflow or mechanical motion, our non-reciprocal system is stable, robust and compact, and it can serve as an excellent platform for topological and non-Hermitian acoustics.

**Results**

The proposed non-reciprocal device is schematically shown in Fig. 1 and it consists of three cavities labeled with A, B and C. The cavities A and C are identical in size and have the same resonance frequency, namely $\omega_C = \omega_A$. Cavity B is detuned by making it smaller, so that it supports a larger resonance frequency $\omega_B$. Between A and C, we introduce a static coupling $k_{AC}$ that introduces a reciprocal transfer phase $\varphi_{static}$. In addition, A and C are interlinked through B with a time-varying coupling coefficient $k_{AB}(t) = \Delta k \cos(\Omega t + \varphi_1)$ and $k_{BC}(t) = \Delta k \cos(\Omega t + \varphi_2)$, where $\Delta k$ is the coupling amplitude, $\Omega = \omega_B - \omega_A$ is the modulation frequency and $\varphi_{1,2}$ are the initial phases. We select the modulation frequency to be the frequency difference between the detuned cavities, realizing efficient frequency conversion and coupling. We define the forward (backward) direction as from A (C) to C (A). To clearly explain the phenomena underlying non-reciprocal transmission, we first set $k_{AC} = 0$. For the forward transmission, the operating frequency of the source is $\omega = \omega_A$, enabling efficient excitation of the resonance in cavity A. According to temporal Floquet theory [32], the converted +1 order harmonic with phase $\varphi_1$ is strictly at $\omega_B$ and can resonantly excite B. Then it will get back to $\omega_A$ with phase $-\varphi_2$ and transfer to C. As a result, the dynamic modulation phases induce an effective magnetic vector potential $\vec{A}$, and the phase for the forward transmission (A to C) is $\varphi_{dynamic}^f = \int_A^B \vec{A} \cdot d\vec{l} + \int_B^C \vec{A} \cdot d\vec{l} = \varphi_1 - \varphi_2$. However, for the backward transmission (C to A), the phase flips sign, namely $\varphi_{dynamic}^b = -\varphi_{dynamic}^f$. Similar to the Aharonov-Bohm effect for electrons [33], here time-reversal symmetry is broken, and the transmission phase is non-reciprocal. When both static and dynamic couplings are considered and the phases satisfy $\varphi_{dynamic}^f \approx \varphi_{static} \approx \pi/2$, there is constructive interference between the two transmitting waves, implying an efficient transfer from A to C. In contrast, when the direction is reversed, $\varphi_{\text{dynamic}}^{\text{b}} \approx -\varphi_{\text{static}}$, and the signal cannot propagate because of destructive interference. As a consequence, we realize frequency-preserving non-reciprocal transmission with large forward transmission and isolation. This mechanism for nonreciprocity is different from, and more generally applicable than synthetic angular momentum bias realized by modulating in time different cavities on site with specific phase patterns [34-36]. In the following, we experimentally verify our proposed scheme and demonstrate non-reciprocal transmission of airborne sound in a practical system.

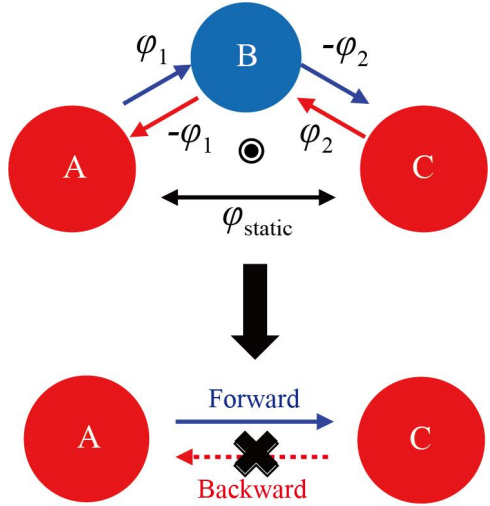

Fig. 1. Schematic diagram of the proposed non-reciprocal system based on three cavities with static and modulated coupling coefficients. The resonance frequency of the cavities is $\omega_C = \omega_A$ and $\omega_B = \Omega + \omega_A$. The static coupling between A and C contributes a reciprocal transmission phase $\varphi_{\text{static}}$. In addition, there are dynamic couplings between A and B (B and C), $k_{AB}(t) = \Delta k \cos(\Omega t + \varphi_1)$ and $k_{BC}(t) = \Delta k \cos(\Omega t + \varphi_2)$. These elements couple the resonant fields to adjacent Floquet harmonics, and the phases induce a synthetic magnetic flux (black point circle) that breaks time-reversal symmetry. Combined static and dynamic couplings induce non-reciprocal power transmission and isolation between A and C.

We used an in-house-designed electric setup to realize tunable coupling among the acoustic cavities. Figure 2(a) shows the photo of the fabricated metallic acoustic cavities and the active circuit consisting of loudspeakers (for output), microphones (for input) and a signal amplifier (with direct current power supply). Sound is detected at resonator A (B) by the microphones, and then coupled to resonator B (A) by loudspeakers after amplification. In this way, the electric coupling between the two cavities can be modulated fast and precisely. We make sure that the realized coupling is linear and bi-directional to avoid spurious nonreciprocal effects that may come from the microphone-to-loudspeaker connection. The cavities are cuboids in shape, and their heights are adjustable to tune the resonant frequencies [37].

In this work, we focus on the first-order resonant mode with a dipole-like mode profile (inset in Fig. 2(b)), which is beneficial to design different coupling effects. By fitting the excitation spectrum of a single cavity, as shown in Fig. 2(b), we get the resonance frequency $\omega_A/2\pi = 1568\text{Hz}$ and intrinsic loss rate $\gamma/2\pi = 4.5\text{Hz}$. The coupling between two such cavities, which is realized through the designed electric setup, is characterized by the splitting of the spectrum into two peaks. This coupling can be modeled with a static Hamiltonian $\mathbf{H} = [\omega_A + \gamma i, k; k, \omega_A + \gamma i]$, where $k$ is the coupling coefficient [38]. According to coupled mode theory, the amplitude $k$ is determined by the distance of the two peaks. Here $|k| = 17$ Hz, whose sign depends on the phase difference between the two cavities, as shown by the simulated field profiles in Fig. 2(c). We define the ratio of pressures between B and A as the transmission $S_{BA}(\omega) = k/(\omega - \omega_A - i\gamma)$ (see Supplementary Material S1 for details [37]). When the microphones and the loudspeakers are connected in-phase, as shown by the red curves, the two cavities are in-phase (out-of-phase) for the higher (lower) frequency range, indicating a positive coupling effect. However, by simply inverting the connections between microphones and loudspeakers, negative coupling is realized due to the flipping of the field parities (see the blue curves and the insets). As a result, our electric setups offer the possibility of designing and tuning the coupling in a way unattainable with existing methods (see Supplementary S3 for comparison [37]) [39-43].

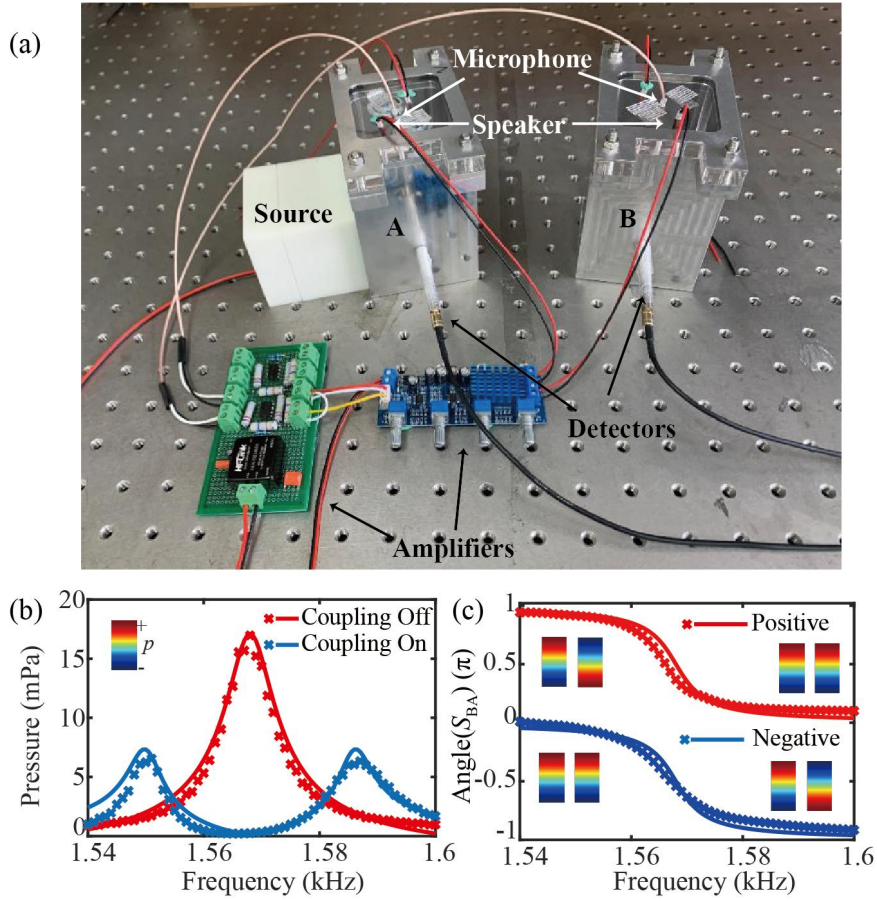

Fig. 2. Tunable coupling with electric control. (a) Experimental design to realize the coupling between two conoid acoustic cavities (labeled A and B). The microphones detect the sound in A (B), and then couple it to B (A) through the loudspeakers after amplification. A sound source is used to excite the system, and other two microphones are inserted into the cavity to extract the pressure information. (b) Measured (crosses) and fitted (solid lines) pressure in A with (blue) and without (red) coupling. (c) The transmission phase arg($S_{BA}$) for positive (red) and negative (blue) coupling.

Next, we design the time-periodic coupling between two cavities, as schematically illustrated in Fig. 3(a). Firstly, we decrease B's height by 4 mm to increase its resonant frequency to $\omega_B/2\pi = 1623$Hz. Two double-pole, double-throw (DPDT) relays are remolded and inserted between the amplifier and the loudspeakers. A square wave voltage signal with $\Omega = \omega_B - \omega_A = 55$Hz is applied to the relays such that the circuit condition changes between in-phase and out-of-phase repeatedly, resulting in a periodic switching between positive and negative coupling. The two cavities only support their first-order resonant modes around 1600Hz. As a result, the frequency conversion process is limited to the first-order Floquet harmonics. It is reasonable to treat the dynamic coupling as $k(t) = \Delta k \cos(\Omega t + \varphi)$, where $\Delta k$ is the effective coupling amplitude and $\varphi$ is the initial phase. Therefore, the Hamiltonian of this time-dependent system is written as

$$\mathbf{H}(t,\varphi) = \begin{bmatrix} \omega_A + i\gamma & \Delta k \cos(\Omega t + \varphi) \\ \Delta k \cos(\Omega t + \varphi) & \omega_B + i\gamma \end{bmatrix}. \tag{1}$$

When A is excited, we simultaneously measure the pressure in these two cavities. The experimental results in Fig. 3(b) show that the acoustic waves are transmitted from A to B with a frequency increase of $\Omega$. To quantitatively evaluate the dynamic coupling effect, temporal coupled model theory (CMT) is employed, for the state function of the system $|\Psi(t)\rangle = [a(t), b(t)]^T$. According to Floquet theory, the signals in the two cavities can be expanded as $a, b(t) = \sum_n a_n, b_n e^{i(\omega + n\Omega)t}$, where $a_n$ and $b_n$ are the time-independent wave amplitude of the $n$-th order Floquet harmonic, and $\omega$ is the excitation frequency. Then the temporal evolution of the system can be described with a Schrödinger-type differential equation

$$-i\frac{d}{dt}|\Psi(t)\rangle = \mathbf{H}(t)|\Psi(t)\rangle + s(t), \tag{2}$$

where $s(t) = [P_{in} e^{i\omega t}, 0]^T$ is the excitation. After some algebra, the coupled-mode equations can be expressed as [44,45]

$$(\omega + n\Omega - \omega_A - i\gamma)a_n - \frac{\Delta k}{2} e^{i\varphi} b_{n+1} - \frac{\Delta k}{2} e^{-i\varphi} b_{n-1} = P_{in} \delta_{n0} \tag{3-a}$$

$$(\omega + n\Omega - \omega_B - i\gamma)b_n - \frac{\Delta k}{2} e^{i\varphi} a_{n+1} - \frac{\Delta k}{2} e^{-i\varphi} a_{n-1} = 0, \tag{3-b}$$

where $\delta_{n0}$ is the Kronecker delta function. From these equations, we see that the cavities are coupled through adjacent harmonics. By truncating the system to finite Floquet harmonics, we solve these coupled equations and obtain the pressure in the two cavities with different coupling strengths. For the case $\omega = \omega_A$, as shown in Fig.

3(c), there is an optimal Δ$k$ that couples the acoustic wave from A to B, because of the trade-off between input coupling and harmonic conversion. By fitting the measured pressure in A, we can estimate the Δ$k$ values. Comparison between Fig. 3(b) and 3(d) suggests that the temporal CMT well explains the dynamic coupling effect.

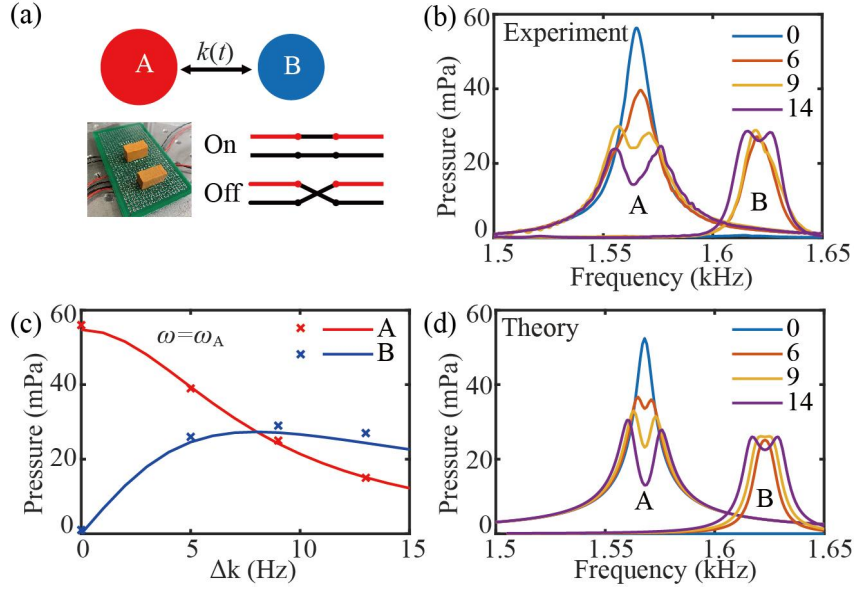

Fig. 3. (a) The upper panel shows the dynamic coupling between two different cavities. The lower panel shows the photograph of the DPDT relays and the on (off) circuit conditions with (without) the controlling voltage signals. (b) Measured pressure in A and B with different coupling strengths Δ$k$. (c) Calculated pressure (solid curves) in A (red) and B (blue) fitting the measured pressure (crosses) for $\omega = \omega_A$. (d) Calculated pressure in A and B for different Δ$k$.

Next, another cavity C is added to form a three-cavity system with non-reciprocal transmission features. As schematically shown in Fig. 1, now the total coupling is composed of a static coupling $k_{AC}$ and two dynamic couplings with initial phase $\varphi_1$ and $\varphi_2$ respectively. For this system, the state function is $|\Psi(t)\rangle = [a(t), b(t), c(t)]^T$ and the temporal Hamiltonian is

$$\mathbf{H}(t,\varphi_1,\varphi_2) = \begin{bmatrix} \omega_A + i\gamma & \Delta k\cos(\Omega t+\varphi_1) & k_{AC} \\ \Delta k\cos(\Omega t+\varphi_1) & \omega_B + i\gamma & \Delta k\cos(\Omega t+\varphi_2) \\ k_{AC} & \Delta k\cos(\Omega t+\varphi_2) & \omega_A + i\gamma \end{bmatrix}. \quad (4)$$

Since the modulation frequency is the difference between cavities, we can perform a transformation to simplify this system: we define a new state function $|\Phi(t)\rangle = M^{-1}|\Psi(t)\rangle$, where

$$M^{-1} = \begin{bmatrix} 1 & 0 & 0 \\ 0 & e^{-i\Omega t} & 0 \\ 0 & 0 & 1 \end{bmatrix}. \tag{5}$$

Using the rotating wave approximation [23], we get a simplified wave function

$$-i\frac{d}{dt}|\Phi(t)\rangle = \begin{bmatrix} \omega_A + i\gamma & \frac{\Delta k}{2}e^{-i\varphi_1} & k_{AC} \\ \frac{\Delta k}{2}e^{i\varphi_1} & \omega_A + i\gamma & \frac{\Delta k}{2}e^{i\varphi_2} \\ k_{AC} & \frac{\Delta k}{2}e^{-i\varphi_2} & \omega_A + i\gamma \end{bmatrix}|\Phi(t)\rangle. \tag{6}$$

Solving these coupled equations, we obtain the transmission coefficients (see Supplementary Material S2 for detailed derivations [37]):

$$s_{CA}(\omega) = \frac{e^{i\Delta\varphi}(\Delta k/2)^2 + (\omega - \omega_A - i\gamma)k_{AC}}{(\omega - \omega_A - i\gamma)^2 - (\Delta k/2)^2}, \tag{7-a}$$

$$s_{AC}(\omega) = \frac{e^{-i\Delta\varphi}(\Delta k/2)^2 + (\omega - \omega_A - i\gamma)k_{AC}}{(\omega - \omega_A - i\gamma)^2 - (\Delta k/2)^2}, \tag{7-b}$$

where $\Delta\varphi = \varphi_1 - \varphi_2$. Clearly, when $\Delta\varphi = \pi/2$ and $k_{AC} = \Delta k^2/4\gamma$, we obtain $s_{AC}(\omega_A) = 0$, which implies that the transmitted waves originated from the static and dynamic coupling effects destructively interfere in the backward direction and the transmission from C to A is suppressed. In contrast, the forward transmission is enhanced, since the two transmitting waves are in phase. In Figs. 4(a-b), we plot the isolation ratio $20\log|s_{CA}(\omega_A)/s_{AC}(\omega_A)|$ and the forward transmission $|s_{CA}(\omega_A)|$ with $\Delta\varphi = \pi/2$. By optimizing the strength of the couplings, we can get efficient and unidirectional power transmission.

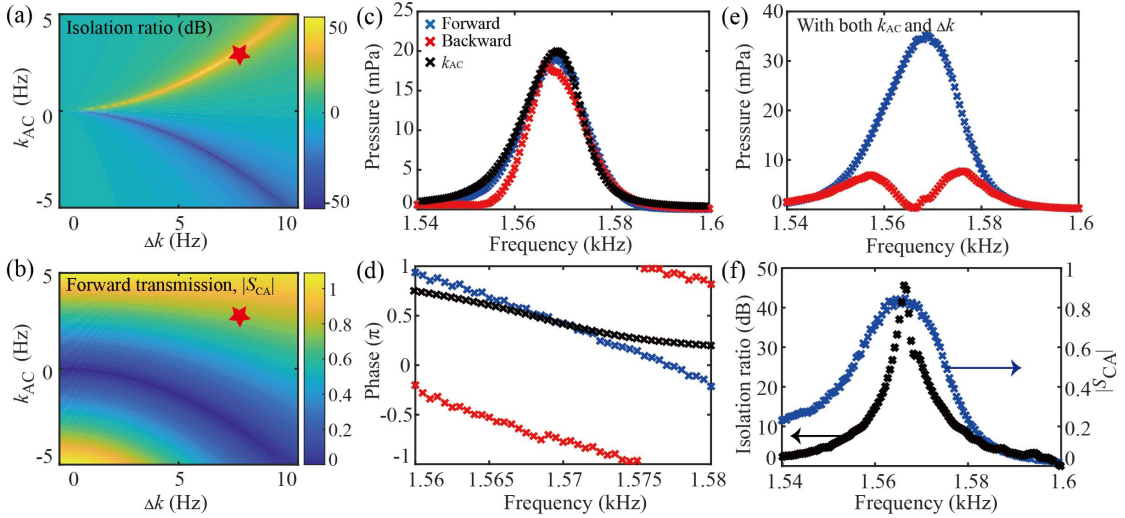

Fig. 4. Non-reciprocal power transmission with $\Delta\varphi = \pi/2$. (a) Calculated isolation ratio and (b) forward transmission for $\omega=\omega_A$. The red stars denote the experimental case with $\Delta k = 7.5$Hz and $k_{AC} = 3$Hz. (c) Measured transmitting pressures and (d) transmitting phases with either $\Delta k$ or $k_{AC}$. (e) Measured transmitting pressures with both $\Delta k$ and $k_{AC}$. (f) Measured isolation ratio and forward transmission.

As an experimental demonstration (red stars in Figs. 4(a-b)), we present the measured amplitudes and phases of the transmitted waves for different coupling conditions, in good agreement with our theoretical predictions (see Supplementary S5 for the comparison [37]). When there is only $\Delta k$, as shown in Figs. 4(c-d), the transmission magnitudes are almost identical for both forward (blue) and backward (red) directions. However, their phases have a difference of π, which is attributed to time-reversal symmetry breaking due to the applied synthetic magnetic field. We note here that the transmission is only non-reciprocal in terms of transmission phase, realizing an acoustic gyrator. On the other hand, when only $k_{AC}$ exists, the transmission amplitudes and phases are identical for both directions (only one direction is shown for simplicity), as expected. When both $\Delta k$ and $k_{AC}$ are present, we realize the expected non-reciprocal power transmission, as seen by the enhanced forward transmission (blue) and nearly-blocked backward transmission (red) near $\omega_A$, as shown in Fig. 4(e). Further quantitative evaluation shows that the maximum isolation ratio reaches 45 dB and the forward transmission is as high as 0.85, as seen in Fig. 4(f). More importantly, it is noteworthy that our mechanism allows further improving the forward transmission by optimizing the coupling strength. As a result, we have realized acoustic nonreciprocity based on temporal modulation of the coupling between cavities, making an important step towards the realization of integrated and compact acoustic isolators. For the cases with $\Delta\varphi = \pi$ or $\Delta\varphi = 0$, the transmission phases for both directions are identical, thus there is no isolation effects (see Supplementary S6-7 for more results [37]).

**Conclusion**

To conclude, we have demonstrated a frequency-preserving non-reciprocal acoustic system based on a mechanism that breaks time-reversal symmetry by utilizing modulation phases to induce a synthetic magnetic field. The measured performance is in good agreement with our theoretical predictions, demonstrating that the isolation reaches 45 dB, while the forward transmission is as high as 0.85. Due to the cascaded conversion processes in the system, the transmitted wave has the original frequency, in stark contrast to previous works in which the transmitting signals are mixed with other harmonics when spatial-temporal modulations or nonlinear media are utilized. Not relying on mechanical motion, our work provides a feasible route to design stable, compact and efficient devices for non-reciprocal sound manipulation, and provides important building blocks to enable acoustic Floquet topological insulators [23,35,46] and non-Hermitian acoustic systems [47,48].

This work was supported by National Key R&D Program of China (Grant No. 2017YFA0303700), the National Natural Science Foundation of China (Grant Nos 11634006, 11374157, and 81127901), a project funded by the Priority Academic Program Development of Jiangsu Higher Education Institutions, the High-Performance Computing Center of Collaborative Innovation Center of Advanced Microstructures, and the Simons Foundation.